\documentstyle[11pt,aas2pp4]{article}
\newcommand{\etal}{{{ et al.}}~}
\newcommand{\eg}{{{e.g.,}}~}
\newcommand{\ie}{{{i.e.,}}~}
\newcommand{\kms}{{{km s$^{-1}$}}~}

\begin{document}

\title{Magnetic Field Evolution in Merging Clusters of Galaxies}

\vspace{1.in}
\author{KURT ROETTIGER}
\affil{Department of Physics and Astronomy\\University of Missouri-Columbia\\Columbia, MO 65211\\email: kroett@hades.physics.missouri.edu}
\author{JAMES M. STONE}
\affil{Department of Astronomy \\ University of Maryland\\ College Park, MD 20742-2421
\\jstone@astro.umd.edu}
\author{JACK O. BURNS}
\affil{Office of Research and Dept. of Physics and Astronomy\\University of Missouri-Columbia\\Columbia, MO 65211\\email: burnsj@.missouri.edu}

\vspace{.5in}

\begin{center}{\bf Accepted for publication in the Astrophysical Journal}
\end{center}




\begin{abstract}

We present initial results from  the first 3-dimensional numerical magnetohydrodynamical (MHD) simulations of 
magnetic field evolution in merging clusters of galaxies.  Within the framework of
idealized initial conditions similar to our previous work, we look at the
gasdynamics and the magnetic field evolution during a major
merger event in order to examine the suggestion that shocks and turbulence generated during a 
cluster/subcluster merger can produce magnetic field amplification and relativistic particle acceleration and, as such, may play a role in the formation and evolution of cluster-wide radio halos.
 The ICM, as represented by the equations of ideal
 MHD, is evolved self-consistently within a changing gravitational potential defined largely by 
the collisionless dark matter component represented by an N-body particle distribution. 
The MHD equations are solved by the Eulerian, finite-difference code, ZEUS. The particles
are evolved by a standard particle-mesh (PM) code.  We find significant evolution of
the magnetic field structure and strength  during
two distinct epochs of the merger evolution. In the first, the field becomes quite filamentary as a result of stretching and compression caused by shocks and bulk flows
during infall, but only minimal amplification occurs. In the second, amplification
of the field occurs more rapidly, particularly in localized regions, as the bulk flow
is replaced by turbulent motions (\ie eddies). The total magnetic field energy is seen
to increase by nearly a factor of three over that seen in a non-merging cluster. In localized regions (associated
with high vorticity), the magnetic energy  can increase by  a factor of 20 or more. A power spectrum analysis of
the magnetic energy shows the amplification is largely confined to scales comparable to and smaller than
the cluster cores indicating that the core dimensions define the injection scale. Although the cluster cores
are numerically well-resolved, we cannot resolve the formation of eddies on scales smaller than approximately
half a core radius. Consequently, the field amplification noted here likely represents a lower limit. We discuss the effects of anomalous resistivity associated with the 
finite numerical resolution of our simulations on the observed field amplification.

\end{abstract}


\section { INTRODUCTION}

Most radio sources in clusters of galaxies are associated with individual
galaxies. There is, however, a class of radio
sources known as radio halos (Jaffe 1977; Hanisch 1980, 1982) which appears to be intrinsic
to the cluster itself rather than any particular galaxy.  Large-scale ($>$0.5 Mpc) radio halos
are rare, diffuse sources which, by their synchrotron emission, demonstrate the existence
of magnetic fields (and a population of relativistic particles) on megaparsec scales. 
Here, we examine numerically the evolution of cluster-wide magnetic fields.

It has been
suggested that the magnetic field is generated by a turbulent dynamo mechanism 
(Jaffe, 1980; Roland 1981; Ruzmaikin, Sokoloff \& Shukurov 1989; Burns \etal 1992; Tribble 1993). Roland (1981)  further suggested that galaxy wakes may be responsible
for generating turbulence throughout the cluster. However, this model has difficulty explaining the radio observations
(Tribble 1991; Goldman \& Raphaeli 1991; DeYoung 1992) because there does not appear to be enough energy in galaxy wakes
to power the radio source. It has been noted that clusters containing radio halos
(such as Coma, A2255, A2256, A2163 etc.) all contain evidence of recent dynamical evolution. That is, they all
exhibit substructure such as non-Gaussian galaxy distributions, multiple X-ray peaks, non-isothermal
temperature distributions, and they lack cooling flows (Edge, Stewart, \& Fabian 1992; Watts 1992; Burns \etal 1995; Burns 1998). This
suggests that cluster mergers are responsible, at least in part, for the formation of radio halos and, by inference,
the growth of magnetic fields on cluster scales (De Young 1992; Tribble 1993).

Cluster mergers are capable of supplying large amounts of kinetic energy, comparable to the thermal 
energy of the cluster, over megaparsec scales. Previous numerical simulations (Roettiger, Burns \& Loken 1993;
Schindler \& M\"uller 1993; Pearce, Thomas \& Couchman 1994; Roettiger, Loken \& Burns 1997) show that mergers generate 
shocks, bulk flows and turbulence within the ICM. The first two of these processes can result in some field
amplification simply through compression. However, it is the turbulence which is the most 
promising source of nonlinear amplification. Provided the turbulence has non-vanishing helicity (  $\langle$ {\bf v}$\cdot \nabla \times$ {\bf v} $\rangle \ne 0$), weak fields
can be amplified exponentially via the well-known $\alpha$-effect (\eg Ruzmaikin \etal 1989). Once the field becomes strong enough
to have a significant back reaction on the flow, the system may
enter an MHD dynamo regime. However, because the full system of MHD equations must be solved to study this regime,
the nature and extent of field amplification which occurs in the MHD dynamo is not well understood.
Of course, in order to account for the synchrotron emission, not only must
there be field amplification, but also the merger induced magnetohydrodynamics must reaccelerate
the relativistic particle population (Pacholczyk \& Scott 1976; Eilek \& Henriksen 1984). Whether the turbulence is responsible
for both, or whether shocks alone are largely responsible for particle acceleration is also not well understood.

In this paper, we begin a study of magnetic field evolution in clusters
of galaxies using three dimensional direct numerical MHD simulations. Here, we
focus in particular on the role of cluster/subcluster mergers, the
resulting gasdynamics (shocks, bulk flows, turbulence), and its role
in the evolution (structure and amplification) of the cluster-wide magnetic field. 
Our numerical method is similar to the cluster merger studies that we have performed in the past,
except in this case, we use an ideal MHD rather than a hydrodynamics code to evolve the ICM.
As in previous simulations, we use idealized, non-cosmological initial conditions consisting
of two isothermal spheres in hydrostatic equilibrium which merge due to the influence of
their mutual gravitational attraction.
This allows us to study in detail the effects of the merger itself separate from
the general cluster formation process thereby optimizing numerical resolution through
an efficient use of the computational volume. It also allows us to control both the
structural and merger parameters of the two clusters, thus allowing for a systematic survey of parameter space.
Since the cluster magnetic field parameters (spatial distribution, strength, length-scale, etc.) are
so poorly constrained at this time, we feel a parameter survey is absolutely essential and will be the subject
of a future study. In
this initial study, we present the results of only a single merger performed at two numerical resolutions, and
a single non-merging cluster.

In \S\ref{num}, we discuss our numerical method. Our initial conditions
are presented in \S\ref{init}.  A discussion of the magnetic field
evolution, including numerical resolution effects, is presented in \S\ref{discuss}. We summarize
our results in \S\ref{summary}.

\section{NUMERICAL METHOD}
\label{num}

The ICM and magnetic fields therein are evolved using
ZEUS (Stone \& Norman 1992a,b), an Eulerian, finite-difference
code which solves, self-consistently, the equations of ideal magnetohydrodynamics
(Jackson 1975). The numerical evolution of the magnetic field components is
performed by the constrained transport (CT) algorithm (Evans \& Hawley 1988) which guarantees
preservation of the divergence-free constraint at all times. The
method of characteristics (MOC) is used for computing the electromotive
force (Hawley \& Stone 1995). An extensive series of MHD test problems have demonstrated that
the MOC-CT method provides for the accurate evolution of all modes of MHD
wave families (Stone \etal 1992). We employ outflow boundary conditions
on the MHD.

The collisionless dark matter is evolved using an N-body code based on a standard 
particle-mesh algorithm (PM, Hockney \& Eastwood 1988). The particles and gas
are evolved on the same grid using the same time step. The time step
is determined by applying the Courant condition simultaneously to both the dark 
matter and the magnetohydrodynamics.
The only interaction between the collisionless particles and the gas is 
gravitational. Since we are modeling an isolated
region, the boundary conditions for Poisson's equation are determined by a multipole expansion (Jackson 1975)
of the total mass distribution (dark matter and gas) contained within the computational grid.  
Particles that leave the grid are
lost to the simulation. Typically, less than a few percent of the particles leave the grid.

The hybrid ZEUS/PM code was parallelized using the Message Passing Interface (MPI; Gropp, Lusk and Skjellum 1994).
These simulations were run on the Cray T3E in the Earth and Space Data Computing Division
of the NASA Goddard Space Flight Center.

In this paper, we report the results of three simulations. In order to assess the
effects of numerical resolution, we have conducted two simulations in which the
resolution differed by a factor of four. The low resolution simulation
was performed on a uniform mesh with dimensions 256 $\times$ 128$^2$ zones. The
high resolution simulation has dimensions which are a factor of two larger (512 $\times$ 256$^2$ zones), but we
gained an additional factor of two in resolution (over the central sub-volume occupied by the cluster
cores and the merger axis) by implementing a non-uniform mesh. The high resolution mesh is uniform from zone 100 to 412 along
the merger axis and from zone 90 to 166 perpendicular to the merger axis. Outside of this central region, we 
gradually increase the zone
sizes, by $\sim$3\%, from one zone to the next radially
outward. The physical dimensions of the computational volume are
scaled to approximately 12 $\times$ 6.5 $\times$ 6.5 Mpc. For the physical scaling describe in \S\ref{init}, our
maximum resolution corresponds to 50 kpc (4.3
zones/core radius) and 12.5 kpc (17.2 zones/core radius), respectively, for the low and high
resolution simulations. The third simulation, also conducted at high resolution, was designed
to assess the numerical dissipation of the magnetic field in the absence of a merger. Here, we simply
evolved a single cluster allowing it to  move across the grid unperturbed. This simulation
provides the baseline for comparison of magnetic field evolution in the merger. During the
single cluster evolution, we note an initial rapid
dissipation of the total magnetic field energy, $\sim$50\%, after which we find variations of less than
20\%.

\section {INITIAL CONDITIONS}
\label{init}
\subsection{Dark Matter and Gas}

Our initial conditions are similar to those used in our previous studies
(\eg Roettiger, Burns \& Loken 1993; Roettiger, Loken \& Burns 1997; Roettiger, Stone \& Mushotzky 1997, 1998).
We begin with two clusters whose gas distributions are  consistent with
observations of relaxed systems. It can be argued that our dark matter profiles which, like the gas profiles,
exhibit a flat core are inconsistent with recent strong lensing observations. 
These observations
seem to imply a central cusp in the dark matter distribution which may be more consistent with the dark
 matter distribution derived by Navarro, Frenk \& White (1997) from cosmological simulations of structure formation.
We do not believe that this difference will significantly alter the merger dynamics. Both clusters in our
simulations have mass 
distributions based on the lowered isothermal King model described
in Binney \& Tremaine (1987). The lowered isothermal King model is a family of
mass distributions characterized by the quantity $\psi/\sigma^2$ which
essentially defines the concentration of matter. As $\psi/\sigma^2$
increases, the core radius ($r_c$) decreases with respect to the
tidal radius ($r_t$). We have chosen a model with  $\psi/\sigma^2$=12 in 
which we have truncated the density distribution at 15$r_c$. Near the half-mass
radius, the total mass density follows a power-law distribution, $\rho 
\sim$ r$^{-\alpha}$, where $\alpha \sim$ 2.6. This model is consistent with
mass distributions produced by cosmological N-body simulations which
show $\alpha \sim$ 2.4 in a high density universe ($\Omega$=1) and $\sim$2.9
in a low density universe ($\Omega$=0.2) (Crone, Evrard, \& Richstone 1994).
Initially the gas distribution (ICM) is in hydrostatic equilibrium within the gravitational
potential defined by both the gas and dark matter components. The gas distribution
is isothermal within the central 6$r_c$. At larger radii, the temperature drops gradually.

Our simulations are conducted in arbitrary units which can be scaled to physically meaningful
parameters by chosing a mass and length scale. Table 1 contains the parameters of the two merging clusters
scaled such that the more massive (or primary) cluster is representative of a rich Coma-like system. 
To date, radio halos have only been identified in relatively rich systems. In line with the physical
scaling used in Table 1, the density peaks are initially separated by 4.6 Mpc. Each cluster was given a small initial
velocity in order to speed up the merger process and thus conserve computational resources.
The final impact velocity ($\sim$2300 \kms) is not  affected significantly by the initial velocity
which has components of 110 \kms parallel to the line of centers and 100 \kms perpendicular to the line of centers. These initial conditions result in a slightly off-axis merger with an impact parameter less than 0.5$r_c$.

\subsection{Magnetic Field}

 Studies of cluster radio halos (Hanisch 1982)
indicate field strengths of order 1 $\mu$G (\eg Coma, Kim \etal 1990; Ensslin \& Biermann 1998; A2255, Burns \etal 1994; A2256, R\"ottgering
\etal 1994) based
on both Faraday rotation measures and equipartition arguments. Unfortunately,
neither of these methods is a direct measure of the field strength, and they usually require
assumptions regarding several poorly constrained parameters (\eg Miley 1980).
 The only direct measurement of magnetic field strengths from synchrotron emission
is via inverse Compton scattering (Felton \& Morrison 1966; Harris \& Grindlay 1979). This method has been used by Bagchi, Pislar \&
Lima Neto (1998) to constrain the magnetic field strength in the diffuse steep spectrum radio source, 0038-096,
in Abell 85. They also find a mean field strength of $\sim$1$\mu$G.
Other analyses of inverse Compton emission have indicated fields of 3-4$\mu$G (Kaneda \etal 1995;
Tashiro \etal 1996), but these appear to be associated with individual radio galaxies and not the cluster itself.
Radio sources in cooling flows have been shown to exhibit large Faraday
rotation measures indicating field strengths in excess of 20$\mu$G (Ge \& Owen 1993; Owen \& Eilek 1998), but, again, these observations
are extremely local and may say more about the cooling flow environment than about the
cluster-wide fields (Soker \& Sarazin 1994). 

 The distribution of magnetic pressure on cluster scales
is similarly uncertain. Deiss \etal (1997) show that the synchrotron halo in Coma,
after subtracting point sources, traces the X-ray surface brightness distribution indicating
that the magnetic pressure gradient is similar to that of the thermal pressure. In addition
to the spatial distribution of the magnetic pressure, we must also be concerned with the
spectral power distribution. The magnetic fields are believed to exist
as tangled flux ropes on a variety of scales (\eg Ruzmaikin \etal 1989). The physical scale on which they are tangled is uncertain. Observations
of polarization in cluster radio sources indicate tangling on scales ranging from a kiloparsec (\eg Feretti \& Giovannini 1997) to tens or even hundreds of kiloparsecs (Kim \etal 1990; Kim, Tribble, \& Kronberg 1991; Feretti \etal 1995) .

Guided as closely by the observational data as possible, we begin by defining a 
magnetic vector potential, A(k)=A$_\circ$k$^{-\alpha}$, where the amplitudes (A$_\circ$) of
each Cartesian coordinate are drawn from
a Gaussian distribution. The vector potential is then transformed via a 3-dimensional
Fast Fourier Transform (FFT) into physical space where it is scaled spatially by the
gas density distribution. Assuming a uniform spherical collapse (likely a gross
over-simplification, \eg Evrard 1990;  Bryan \etal 1994; among many others) and flux freezing, it can be shown that the magnetic field will scale
as $\rho_{gas}^{2/3}$. Next,
we scale the amplitude of the magnetic pressure such that the maximum magnetic energy density is equal
to 1\% of the local thermal pressure. This leads to a mean field of 0.07 $\mu$G within
the central 2$r_c$ of the primary cluster. 
Finally, we initialize a tangled divergence-free magnetic field from {\bf A} via ${\bf B}=\nabla \times {\bf A}$.  
This results in {\bf B$^2$} $\propto$ k$^{2(1-\alpha)}$. Here, we adopt $\alpha=5/3$. The premerger profile shapes (normalized at the core radius) can be seen in Figure \ref{prof}.

\section {DISCUSSION}
\label{discuss}

\subsection{Magnetic Field Evolution}
\label{mfev}

 A more detailed
description of the hydrodynamical and N-body evolution of merging clusters of galaxies can be found in Roettiger, Burns, \& Loken
(1996); Roettiger, Loken, \& Burns (1997); Roettiger, Stone \& Mushotzky (1997, 1998).
In fact, since the initial fields are weak, the hydrodynamics is largely the same.
Here, we focus on the magnetic field evolution. 

Figure \ref{devol} shows the evolution of gas density (Column 1), gas temperature (Column 2) and 
magnetic pressure (B$^2$, Column 3) at
four epochs (Rows 1-4) during the merger. The epochs depicted here correspond to 0.0, 1.3, 3.5, and 5.0 Gyrs
after the time of closest approach. Each panel represents a 2-dimensional slice taken through the core of
the merger remnant and in the plane of the merger. The slice dimensions (3.75 $\times$ 3.75 Mpc) are only a small fraction of the simulation volume. 

In Figure \ref{devol}a,b,c, the smaller of the two clusters (hereafter the subcluster) is seen impinging
on the core of the larger cluster (or primary) from the right hand side and moving toward the left.
Since the cluster gravitational potential is quite steep and since gas is continually 
being stripped from the subcluster, a gradient in the gas velocity appears across the subcluster.
The result is a stretching, and consequently, a small amplification of the magnetic field
which produces long radial filaments, particularly in the subcluster wake (Figure \ref{devol}c).
 As the
merging cores become coincident, the gravitational potential reaches an extreme minimum
drawing in gas from all directions which further enhances the radial filamentary structure.
A shock forms along the leading edge of the subcluster resulting in compression and
amplification of the magnetic field along the shock front, which is evident by the arc of hot gas visible 
in the temperature data, Figure \ref{devol}b.

At 1.3 Gyrs (Figure \ref{devol}d,e,f), 
the bow shock has propagated off the left hand side of the frame having left behind a sheet of enhanced magnetic field 
which had been compressed along its leading edge (Figure \ref{devol}f).
In the subcluster wake, the magnetic field is drawn out of the primary cluster into long filaments. 
It is this type of magnetic field distribution which may explain the south-west extension of the Coma cluster radio halo
(Deiss \etal 1997). If so, this would
tend to support the assertion that the NGC 4839 group has already fallen through the core of Coma
and is dragging magnetic field with it (Burns \etal 1994).

Once the dark matter component of the subcluster exits the primary core, the gravitational potential minimum quickly returns
to near premerger values. This causes a rapid
expulsion of the gas that was drawn in during the merger. The expelled gas interacts with
residual infalling gas from the subcluster creating a second shock that propagates upstream (to the right)
resulting in further compression and amplification of the field. Morphologically, there is considerable
similarity between the two shocks and magnetic field structures (Figure \ref{devol}f) discussed here and the diffuse radio halo 
 observed in A3667 (R\"ottgering \etal 1997). 
We address the possible physical connection between the merger induced magnetohydrodynamics seen
here and the A3667 radio halo in a future
paper (Roettiger, Burns, \& Stone 1999).

During the early stages of the merger, the cluster gasdynamics are dominated by large scale bulk flows
and large eddies in the wake of the subcluster. Between 1.3 Gyrs and 2.5 Gyrs, the subcluster dark
matter remnant  passes through the cluster core three times. Although each passage is less extreme than the
one before, each contributes to the breakdown of the bulk flows and to the randomization of the gas velocities.
It is during this phase that the most extreme amplification of local field energy occurs (26$\times$ over a non-merging cluster). 
After 2.5 Gyrs, the magnetic field extrema begin to decay with time while the total magnetic energy continues to grow.  
The breakdown of the long filamentary structures is seen to begin in the core (Figure \ref{devol}i) and proceed radially outward
as the cluster relaxes (Figure \ref{devol}l).

Figure \ref{btotevol} (solid line) shows the increase in total magnetic energy within the
entire computational volume relative to the non-merging single cluster. Note the slight increase ($\sim$30\%)
during the first core passage ($t=$0). This
is a result of the initial compression which, although it approaches the strong shock conditions, is 
very local and consequently does not greatly effect the total magnetic energy within the
larger volume.  From core passage to the
end of the merger simulation, the magnetic energy grows essentially linearly with time increasing by 
nearly a factor of 2.75 after 4 Gyrs.  We find that the fractional field amplification
within the cluster core is comparable to that seen in the much larger volume. Of course the initial
compression at $t=$0 makes a much larger relative contribution to the smaller volume. There is then a decline
in the magnetic energy as the field is drawn out of the cube only to return with the subcluster remnant
after about 1.3 Gyrs.
The dashed line in Figure \ref{btotevol} shows the growth of the total magnetic field energy relative to
the thermal energy also normalized to the single cluster evolution. After passage of the initial shock ($t=$0),
the relative increase of magnetic energy to thermal energy tracks the relative increase in the magnetic energy
quite well (\ie the solid and dashed lines in Figure \ref{btotevol} are essentially parallel after $t=$0).
Figure \ref{totenergy} shows the evolution of the total thermal, kinetic and magnetic energy densities
within 400 kpc of the gravitational potential minimum. These quantities are not normalized and have been
scaled to the physical dimensions in Table 1. Here we note the fractionally large increase in both thermal and
kinetic energy, particularly at the time of core passage ($t=$0). The increase in magnetic energy
is far more modest, indicating that a very small fraction of the merger energy goes into field amplification.

It is important to note that although the total magnetic energy has increased by only a factor
of $\sim$2.75, local amplification can be considerably greater. At $\sim$2 Gyrs after the initial core
passage, the maximum magnetic pressure has increased by a factor of 12 over premerger values and
by a factor of 26 over the non-merging cluster at the same epoch. The peak maxima in magnetic pressure
occur between 1.3 and 2.5 Gyrs. As mentioned above, at 1.3 Gyrs, the initial bulk flows begin
to breakdown with the second passage of the subcluster remnant, and at 2.5 Gyrs, stirring by the
subcluster remnant largely ceases. The peak magnetic pressure after 5 Gyrs
is still a factor of three greater than both  premerger values and the non-merging cluster at the same
epoch.

Figure \ref{prof6} shows the azimuthally averaged and normalized profiles of gas density, thermal pressure
and magnetic 
pressure at 5 Gyr after closest approach. Comparison with the initial profiles in Fig. \ref{prof},
shows relatively little evolution in the profile shapes. At this late stage of the merger the cluster properties have
had time to equilibrate, and profiles tend to smooth out localized substructure. Numerical reconnection and dissipation of the magnetic field has proceeded
in the core on scales less than 4-5 zones (50-60 kpc). We also note an increase in the thermal pressure
of the core relative to the outer regions of the cluster. This is a common feature of the merger simulations
(Roettiger \etal 1997). Much of the kinetic energy of the merger is dissipated within the core resulting in
a rise in temperature accompanied by an expansion of the gas distribution.

Another way to look at the magnetic field evolution is through an analysis of the spectral power distribution.
Figure \ref{bpower} shows the ratio of the pre- to post-merger magnetic field power spectrum at three post-merger
epochs. At 1.3 Gyrs, before turbulent gas motions have 
replaced the bulk flows, power has increased by less than a factor of 10 on resolved ($>$4-5 zones) scales.
At later times power has increased by nearly a factor of 20 on the scale of $\sim$4 zones and greater than
a factor of 10 on scales less than 6-7 zones. Again, this is consistent with our factor of three increase
in total magnetic energy since relatively little power resided in these small scales initially.

Field amplification occurs on scales less than $\sim$8-10 zones. This is comparable to the
diameter  of the subcluster's gas core. Although numerical resolution is likely still playing a role, it is
reasonable to expect that the largest scale on which turbulent eddies form and have sufficient
time to turn over and amplify the field would be comparable
to the subcluster core dimensions. The important dimension is the gas core diameter, because it is the hydrodynamical interaction between the clusters that is ultimately responsible for
the field amplification, and the gas core supplies the greatest hydrodynamical impact. Subcluster gas
at larger radii is much less dynamically significant (\ie of much lower density and velocity) and is largely stripped by the time the cluster cores interact.
Unfortunately, even at 8-10 zones we may still be uncomfortably close
to the resolution limitations. Given greater resolution, we would expect the eventual formation
of smaller eddies which could potentially result in even greater field amplification
on small scales. Small eddies turn over more rapidly than large eddies and will thus amplify
the field to a greater extent over a given time scale.

\subsection{Resolution Study}
\label{res}

In order to study the effects of numerical resolution, we have conducted two simulations that
differed in effective resolution by a factor of four. In the low resolution simulation,
the primary cluster core is resolved by 8-9 zones while in the high resolution simulation (the results
of which were discussed in \S\ref{mfev}),
the core is resolved by 34-35 zones. In ZEUS, an artificial viscosity is used to thermalize
kinetic energy in shocks; this viscosity smooths shocks over 4-5 zones. Similarly, the magnetic field
resolution, (\ie the scale over which numerical reconnection occurs) is $\sim$4 zones.
In both simulations, we see the  formation of long filaments  during the early stages of the merger. 
The width of these filaments appear to be determined by the effective resolution.

In the low resolution simulation, the field evolution largely ceases after the formation of the
filaments. The filaments are advected about the cluster by residual gas motions, but no further
amplification occurs. This is in contrast to the high resolution simulation which shows a factor of 2.75
increase in the total magnetic energy after 4 Gyrs. It appears that the lack of numerical resolution
prevents the development of turbulent eddies on meaningful scales.
That is, the injection scale for turbulence in these mergers is comparable to the cluster core dimensions.
Since the subcluster core is only marginally resolved in the low resolution simulation, turbulent eddies
are not resolved well enough to form nor if they form do they persist long enough for amplification of 
the magnetic field to occur.
Figure \ref{bpower_res} shows the ratio of the pre- to post-merger power spectra for both resolutions
at 3.4 Gyrs after core passage. The low resolution simulation shows no significant amplification on
scales greater than 4-5 zones, and the amplification that is evident is considerably less than
it is in the high resolution simulation. At this time, there is no convergence in the magnetic field
evolution. We speculate that given higher resolution we could follow the cascade of eddies to smaller
and smaller scales expecting greater field amplification as the turn over timescale for the eddies
decreases. Unfortunately, at this time, a significant increase in resolution over our high resolution
simulation is not technically feasible given the limitations of currently available computational
platforms.

Numerical dissipation will have two effects on the MHD of the mergers: 1) numerical viscosity will 
artificially truncate  the spectrum of turbulent eddies on scales of a few grid zones, and 2) numerical
resistivity will allow for anomalous dissipation of the field on similar scales. The actual conductivity and viscosity of the ICM may
have significantly different characteristic scales, so that the magnetic Prandtl number (the
ratio of the coefficient of viscosity and resistivity) may be much different
than one. Thus, studying the degree of amplification of the magnetic field as the magnetic
Prandtl number is varied in the simulations is of great interest. However, 
since this requires a large dynamic range from core radius to the dissipation scales (the latter
of which must be resolved in order to capture the turbulent eddies), such studies appear to be
well beyond the range of current computational resources. Our largest simulation is far from having the requisite
dynamic range in the remnant core for such studies.

\section {SUMMARY}
\label{summary}

We have presented the initial results from 3-dimensional numerical
MHD/N-body simulations of merging clusters of galaxies. We find that
cluster mergers can dramatically alter the local strength and
structure of cluster-wide magnetic fields. Early in the merger,
there is a filamentation of the field caused by tidal forces acting
on the ICM component of both clusters. The steepness of the gravitational potential causes a
gradient in the gas velocity which results in the stretching
and filamentation of the field.  We also find compression and amplification
of the field along a shock located at the leading edge of the
subcluster. Once formed, this structure propagates with the shock
through the core of the cluster and out the other side. A similar
feature is formed on the upstream side of the merger when gas that is expelled
from the rapidly varying gravitational potential interacts with residual
infall from the subcluster. The result is two shocks, which through compression form
magnetic field structures on
opposing sides of the merger remnant core. The strong shocks associated with these
structures may be the source of energetic particles to power the synchrotron
emission. These structures
are similar in morphology to the radio halo in A3667 (R\"ottgering \etal 1997).
Similarly, we find an extension of the magnetic field distribution along the
merger axis toward the subcluster which is reminiscent of the southwest extension 
of the radio halo in Coma (Deiss \etal 1997).

In the early stages of the merger, amplification of the field is
limited to that produced by compression in the bow shock and by stretching resulting
from infall. It is only after the merger induced bulk flows breakdown into turbulent
gas motions that  significant local  field amplification takes place.
Since we find basically two epochs of magnetic field evolution,
we suggest that there are two epochs of radio halo formation. The first, represented by the
halos in A2255 (Burns \etal 1995) and A2256 (R\"ottgering \etal 1994), are in
the early stages of a merger and result largely from field compression associated with the
initial infall. The second, represented by the halo in Coma (Kim \etal 1990), is in a later
stage of the merger dominated more by turbulent gas motions.

We find that as a result of the merger the total magnetic energy within the
computational volume rises steadily by
nearly a factor of three during the 5 Gyrs after core passage. On smaller scales
we find that the magnetic field energy increases by greater than a factor of 10 to 20. Of course, we have modeled
only a single merger. It is likely 
that massive clusters will undergo several major mergers during their life time 
(every 2-4 Gyrs; Edge \etal 1992) and that
each successive merger will further amplify the field.  Also, it is likely that the ICM in massive 
clusters is being stirred almost continuously by  galaxy wakes (Roland 1981; De Young 1992; Merrifield 1998) 
and by residual infall including many
lesser mergers (Norman \& Bryan 1998).  The increase in the magnetic energy is only
about $10^{-5}$ of the kinetic energy imparted by the merger event. The remaining energy goes
into  residual gas motions and heating of the ICM. This value may be limited by numerical
resolution effects.

Our comparison of the premerger and post-merger power spectra of the magnetic 
field distribution show that the greatest fractional amplification 
of the field is on relatively small scales.  There are several reasons for
this. First,
the most extreme amplification is associated with localized gasdynamics
 such as shocks and small scale resolved eddies which have the fastest
turn-over time and therefore the shortest amplification time.  Second, the
scale of the merger induced turbulence is strongest on  scales smaller
than the perturbation, in this case, the core diameter of the impinging subcluster
(17 zones in the high resolution simulation).

Resolution effects
are always important in a study of this type. Observations indicate that magnetic fields, like
turbulence, have power on a range of physical scales. It is impossible,
at this time, to simulate all scales accurately.
We note a couple of significant resolution effects. First, the width 
of the filaments formed during
the early stages of the merger appears to be determined
by the numerical resolution. This is also true of the magnetic sheets that form along
shock structures. Second, and most importantly, in order to see
significant amplification, turbulent eddies must be resolved. In these
simulations eddies are resolved at 4-8 zones across.  Finally, if the small scales (1 kpc; Feretti \etal 1995) for field tangling deduced from the radio observations are
correct, then we could be missing a very important dynamical scale in these simulations.
The dynamic range required to resolve the full spectrum of turbulent eddies on scales below the
core radius, and yet still model the global
merger, presents an enormous challenge to future numerical studies.
Future work will include a survey of magnetic field
initial conditions (\eg length scales, pressure distribution),  and the modeling of 
specific clusters containing radio halos (\eg A3667; Roettiger \etal 1999).

\bigskip

We thank the anonymous referee for many useful suggestions that have improved the presentation
of this material. We thank the Earth and Space Data Computing Division at the NASA Goddard Space Flight Center
for use of the Cray T3E supercomputer. KR would also like to thank the National Research
Council for financial support during the early stages of this work. JB acknowledges support
from NSF grant AST-9896039.
\newpage

\section*{ REFERENCES}
 \everypar=
   {\hangafter=1 \hangindent=.5in}

{

Bagchi, J., Pislar, V., \& Lima Neto, G. B. 1998, MNRAS, submitted (astro-ph/9803020)

Binney, J, \& Tremaine, S. 1987, Galactic Dynamics, (Princeton: Princeton University Press)

Bryan, G. L., Klypin, A., Loken, C., Norman, M. L., \& Burns, J. O. 1994, ApJ, 437, 5L

Burns, J. O. Sulkenen, M. E., Gisler, G. R., \& Perley, R. A. 1992, ApJ, 338, L49

Burns, J. O., Roettiger, K., Pinkney, J., Perley, R. A., Owen, F. N., \& Voges, W. 1995, ApJ, 446, 583

Burns, J. O., Roettiger, K., Ledlow, M. \& Klypin, A. 1994, ApJ, 427, L87

Crone, M., Evrard, A., \& Richstone, D. 1994, ApJ, 434, 402

DeYoung, D. S. 1992, ApJ, 386, 464

Deiss, B. M., Reich, W., Lesch, H., \& Wielebinski, R. 1997, A\&A, 321, 55

Edge, A. C., Stewart, G. C., \& Fabian, A. C. 1992, MNRAS, 258, 177

Eilek, J. A. \& Henriksen, R. N. 1984, ApJ, 277, 820

Ensslin, T. A. \& Biermann, P. L. 1998, A\&A, astro-ph/9709232

Evrard, A. E. 1990, ApJ, 363, 349

Evans, C. R. \& Hawley, J. F. 1988, ApJ, 332, 659

Felton, J. E. \& Morrison, P . 1966, ApJ, 146, 686

Feretti, L., Dallacasa, D., Giovannini, G., \& Tagliani 1995, A\&A 302, 680

Feretti, L., \& Giovannini, G. 1997, astro-ph/9709294

Ge, J. P., \& Owen, F. N. 1993, AJ, 105, 778

Gropp, W., Lusk, E., \& Skjellum, A. 1994, Using MPI: Portable Parallel Programming with the Message-Passing Interface, (Cambridge: MIT Press)

Hanisch, R. J. 1980, AJ 85, 1565

Hanisch, R. J. 1982, A\&A 116, 137

Harris, D. E. \& Grindlay, J. E. 1979, MNRAS, 188, 25

Hawley, J. F., \& Stone, J. M. 1995, Comp. Phys. Comm., 89, 127

Hockney, R., \& Eastwood, J. 1988, Computer Simulation Using Particles, (Philadelphia: IOP)

Jackson, J. D. 1975, Classical Electrodynamics, (New York: Wiley)

Jaffe, W. J. 1977, ApJ, 212, 1

Jaffe, W. J. 1980, ApJ, 241, 925

Kaneda, H. \etal 1995, ApJ, 453, 13L

Kim, K.-T., Kronberg P. P., Dewdney, P. E., Landecker, T. L., 1990, ApJ, 355, 29

Kim, K.-T., Tribble, P. C., \& Kronberg, P. P. 1991, ApJ, 379, 80

Merrifield, M. R. 1998, MNRAS, 294, 347

Miley, G, 1980, ARAA, 18, 165

Navarro, J. F., Frenk, C. S., White , S. D. M. 1997, ApJ, 490, 493

Norman, M. L. \& Bryan, G. 1998, Ringberg Workshop on M87, eds. K. Meisenheimer \& H.-J. R\"oser, Springer Lecture Notes in Physics, in press (astro-ph/9802335)

Owen, F. N. \& Eilek, J. A. 1998, ApJ, 493, 73

Pacholczyk, A. G. \& Scott, J. S. 1976, ApJ, 203, 313

Pearce, F. R., Thomas, P. A., \& Couchman, H. M. P. 1994, MNRAS, 268, 953

R\"ottgering, H., Snellen, I., Miley, G., de Jong, J. P., Hanisch, R., \& Perley, R. A. 1994, ApJ, 436, 654

R\"ottgering, H., Wieringa, M. H., Hunstead, R. W., \& Ekers, R. D. 1997, MNRAS, 290, 577

Roettiger, K., Burns, J. O., \& Loken, C 1993, ApJ, 407, 53L

Roettiger, K., Burns, J. O., \& Loken, C 1996, ApJ, 473, 651

Roettiger, K., Loken, C., \& Burns, J. O. 1997, ApJS, 109, 307

Roettiger, K., Stone, J. M., \& Mushotzky, R. F. 1997, ApJ, 482, 588

Roettiger, K., Stone, J. M., \& Mushotzky, R. F. 1998, ApJ, 493, 62

Roettiger, K., Burns, J. O., \& Stone, J. M. 1999, ApJ, submitted.

Roland, J. 1981, A\&A, 93, 407

Ruzmaikin, A., Sokoloff A., \& Shukurov, A. 1989, MNRAS, 241, 1

Sarazin, C. 1986, Rev. Mod. Phys., 58, 1

Schindler, S. \& M\"uller, E. 1993, A\&A, 272, 137

Soker, N., \& Sarazin, C. 1990, ApJ, 348, 73

Stone, J. M. \& Norman, M. 1992a, ApJS, 80, 753

Stone, J. M. \& Norman, M. 1992b, ApJS, 80, 791

Stone, J. M., Hawley, J, F., Evans, C. R., \& Norman, M. L. 1992, ApJ, 388, 415

Tashiro, M. 1996, ``Xray Imaging and Spectroscopy of Cosmic Hot Plasmas",
eds. F. Makino, K. Mitsuda, (Tokyo: Universal Academy Press, Inc.), 83

Tribble, P. C. 1991, MNRAS, 253, 147

Tribble, P. C. 1993, MNRAS, 263, 31

}

\newpage

\begin{center}{\bf FIGURE CAPTIONS}
\end{center}

\begin{figure}[htbp]
\centering \leavevmode
\epsfxsize=1.\textwidth \epsfbox{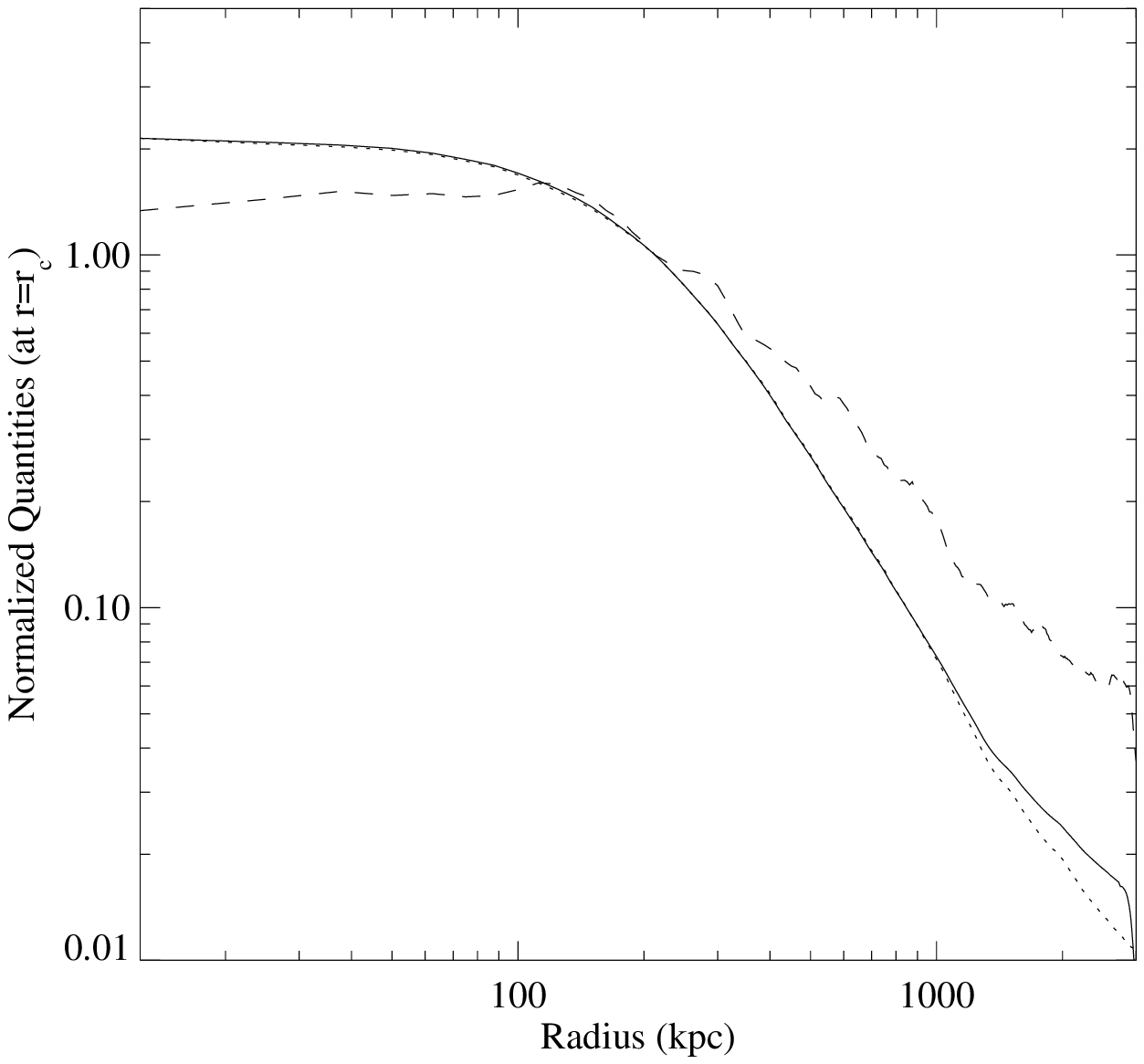}
\caption[]
{ }
\label{prof}
\end{figure}

 {\bf Fig. \ref{prof}} Premerger gas density (solid line), thermal pressure (dotted line) and magnetic
pressure (dashed line) profiles of the primary cluster normalized to their values at the core radius.
The central dip in the magnetic pressure is the result of numerical reconnection of the field. At this time, the clusters have been evolved approximately 3 Gyrs.
The merger will occur in approximately 1.5 Gyrs.

\begin{figure}[htbp]
\centering \leavevmode
\epsfxsize=0.9\textwidth \epsfbox{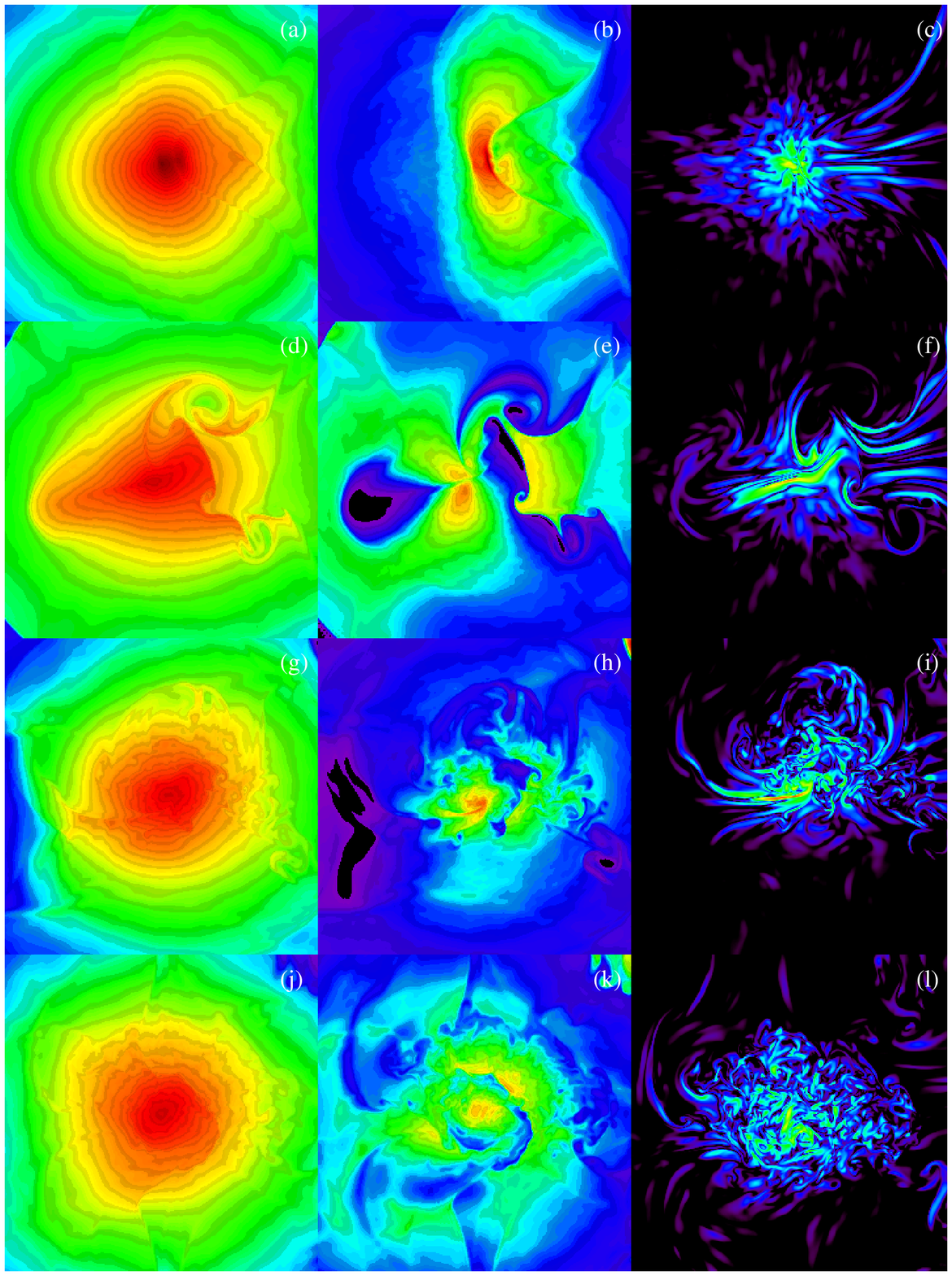}
\caption[]
{ }
\label{devol}
\end{figure}

{\bf Fig. \ref{devol}} The evolution of log gas density (column 1), gas temperature (column 2), and log
magnetic pressure (column 3) in two-dimensional slices taken through the cluster core in the plane
of the merger. Each row represents a different epoch. Row 1 is $t$=0, the time of core coincidence.
Rows 2, 3, and 4 represent $t$=1.3, 3.4, and 5.0 Gyrs, respectively. In each quantity, black represents
low values and red represents high values. The color transfer function
is consistent from one panel to the next within a given physical quantity.
Each panel is 3.75$\times$3.75 Mpc. The range of values in
each quantity are: gas density, 6$\times$10$^{-6}$ to 9.2$\times$10$^{-3}$ cm$^{-3}$; gas temperature,
2.6 to 23.1 keV; and magnetic pressure, 1.24$\times$10$^{-21}$ to 5.9$\times$10$^{-13}$ dynes cm$^{-2}$.

\begin{figure}[htbp]
\centering \leavevmode
\epsfxsize=1.\textwidth \epsfbox{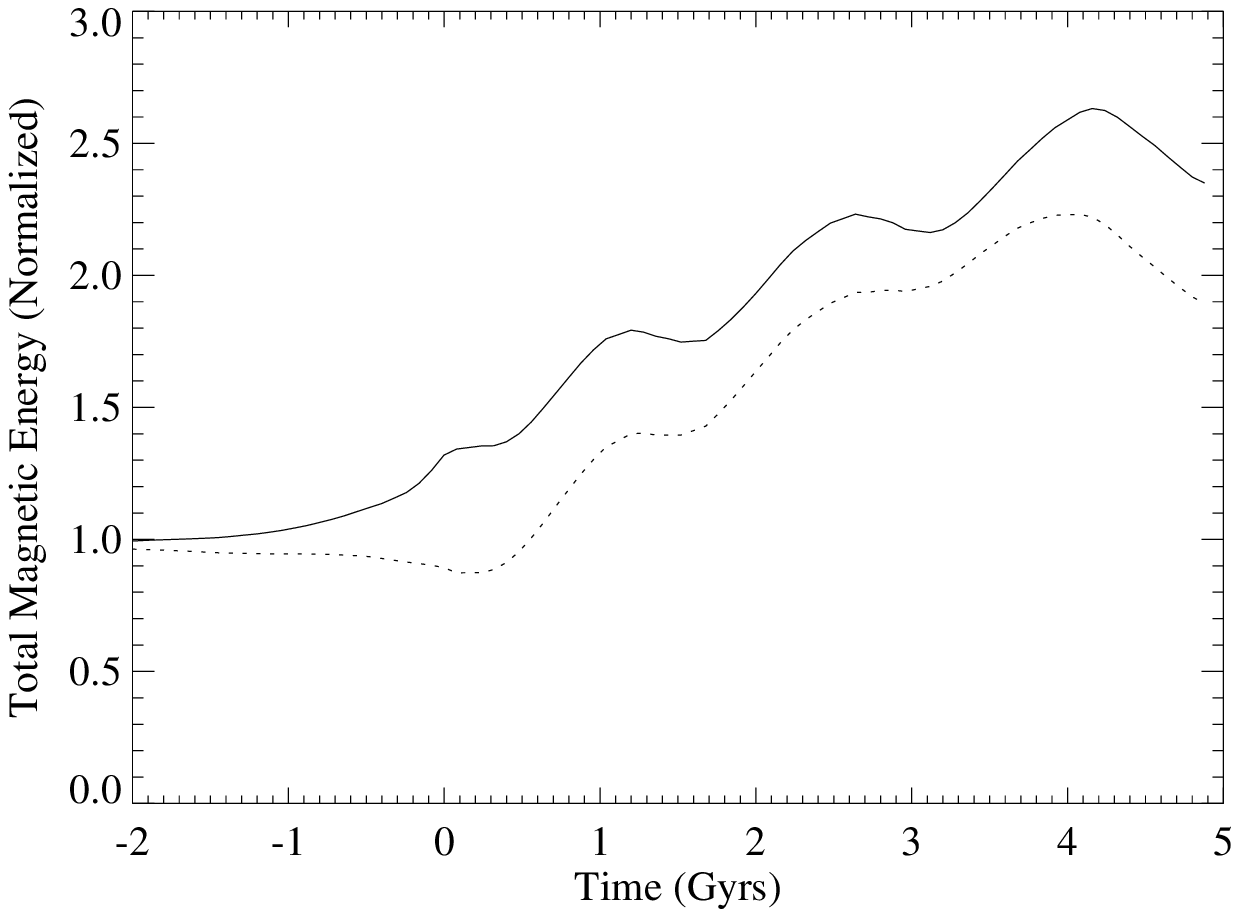}
\caption[]
{ }
\label{btotevol}
\end{figure}

{\bf Fig. \ref{btotevol}} The total magnetic energy within
the entire computational volume (solid line) and  the evolution of 
the magnetic to thermal energy ratio (within the entire volume), also normalized to a non-merging cluster (dashed line).
Time is given relative to the time of closest approach, or core passage.

\begin{figure}[htbp]
\centering \leavevmode
\epsfxsize=1.\textwidth \epsfbox{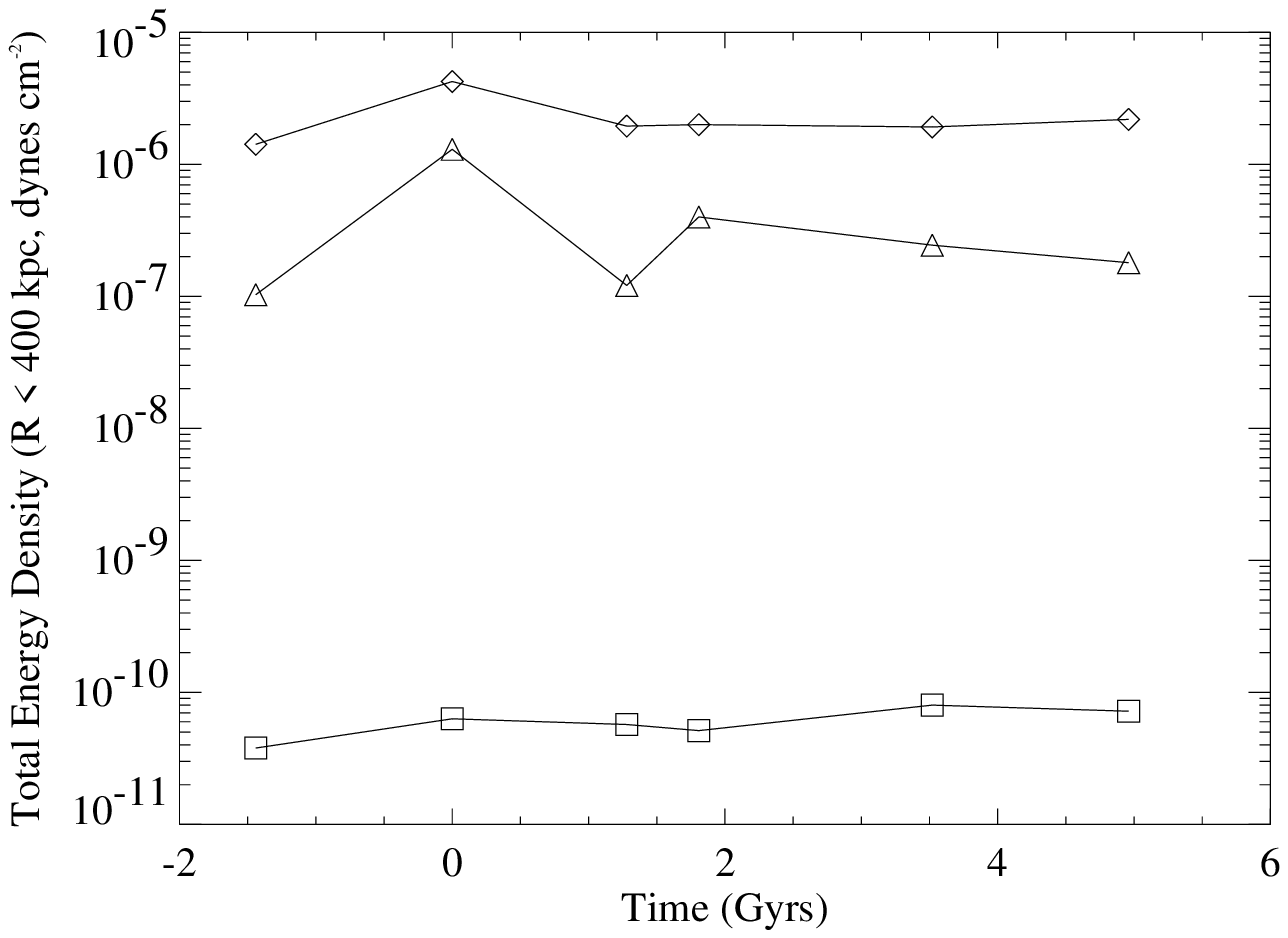}
\caption[]
{ }
\label{totenergy}
\end{figure}

{\bf Fig. \ref{totenergy}} The total thermal (open diamonds), kinetic (open
triangles) and magnetic (open squares)  energy densities within an 800 kpc cube  centered on the gravitational potential minimum.
The quantities are scaled according to the physical dimensions in Table 1.
Time is given relative to the time of closest approach, or core passage.

\begin{figure}[htbp]
\centering \leavevmode
\epsfxsize=1.\textwidth \epsfbox{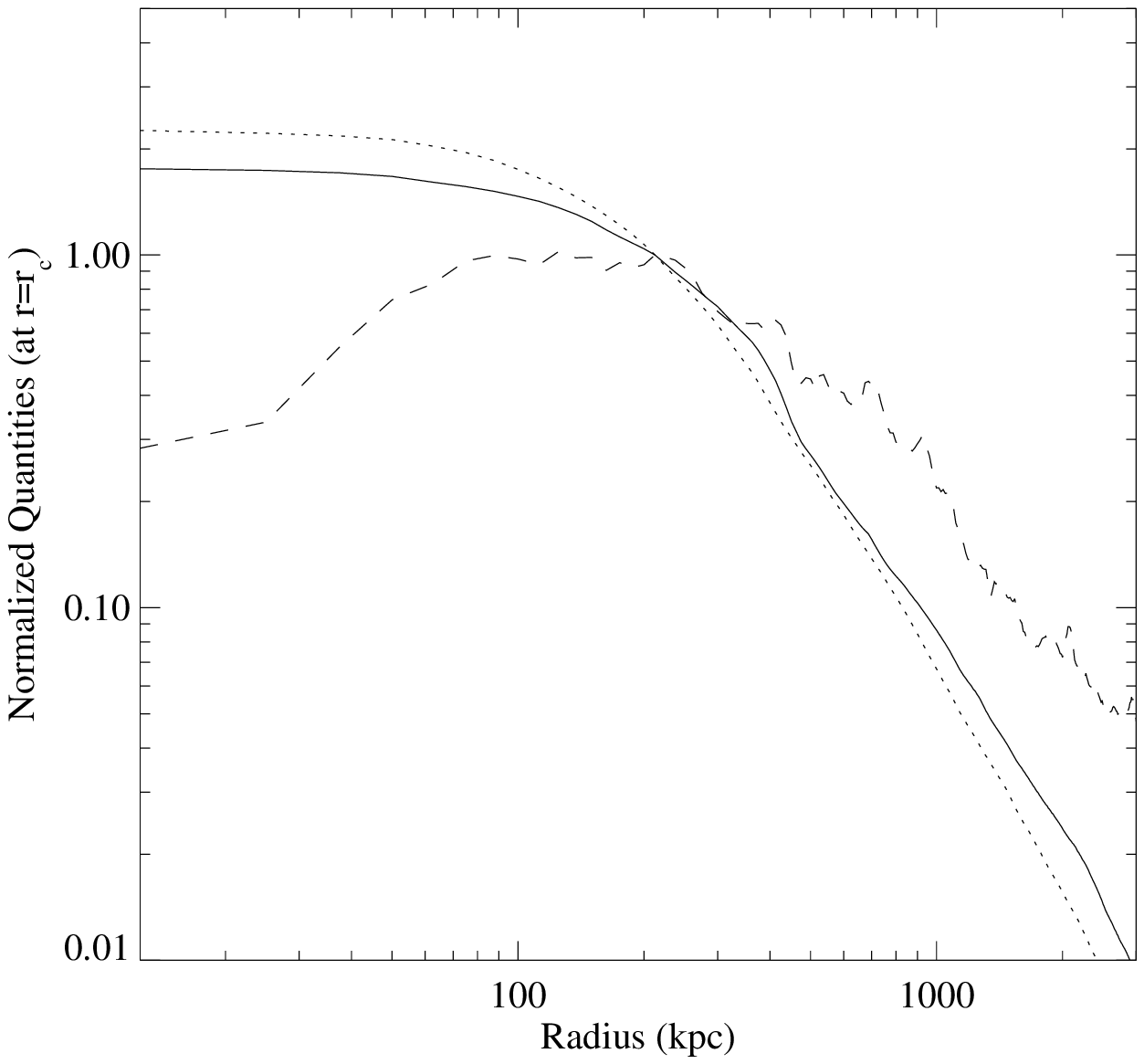}
\caption[]
{ }
\label{prof6}
\end{figure}

{\bf Fig. \ref{prof6}} The gas density (solid line), thermal pressure (dotted line) and magnetic
pressure (dashed line) profiles of the merger remnant normalized to their values at the core radius.
The central dip in the magnetic pressure is the result of numerical reconnection of the field. At this time, the clusters have been evolved approximately 5 Gyrs
since core passage.

\begin{figure}[htbp]
\centering \leavevmode
\epsfxsize=1.\textwidth \epsfbox{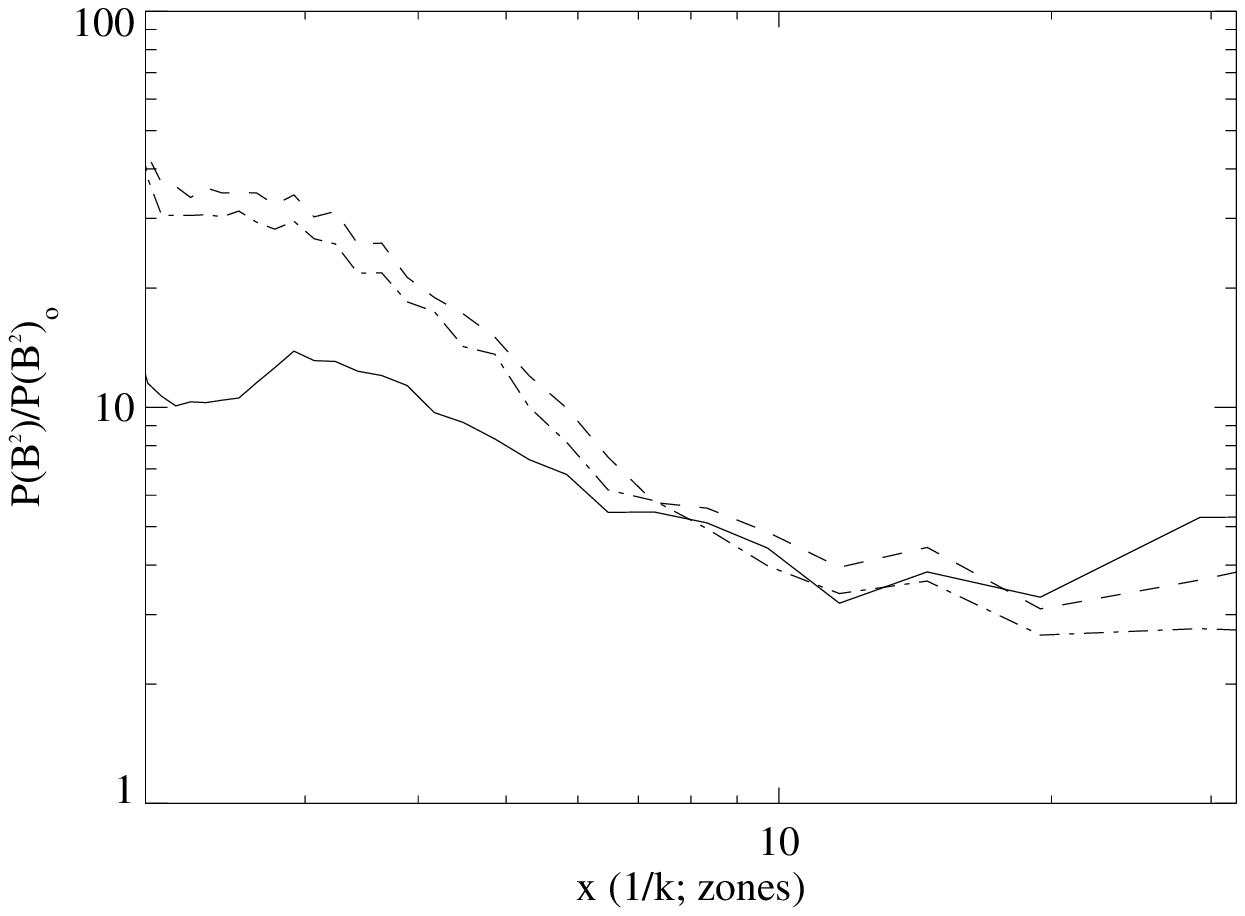}
\caption[]
{ }
\label{bpower}
\end{figure}

{\bf Fig. \ref{bpower}} The ratio of  pre- to post-merger magnetic energy power spectra at three different
epochs, 1.3 Gyr (solid line), 3.4 Gyrs (dashed line), and 5.0 Gyrs (dash-dot line), for the high
resolution merger. This analysis was performed on an 800 kpc cube (64$^3$) centered on the gravitational
potential minimum. The scale for significant amplification appears to be less than 8 to 10 zones.

\begin{figure}[htbp]
\centering \leavevmode
\epsfxsize=1.\textwidth \epsfbox{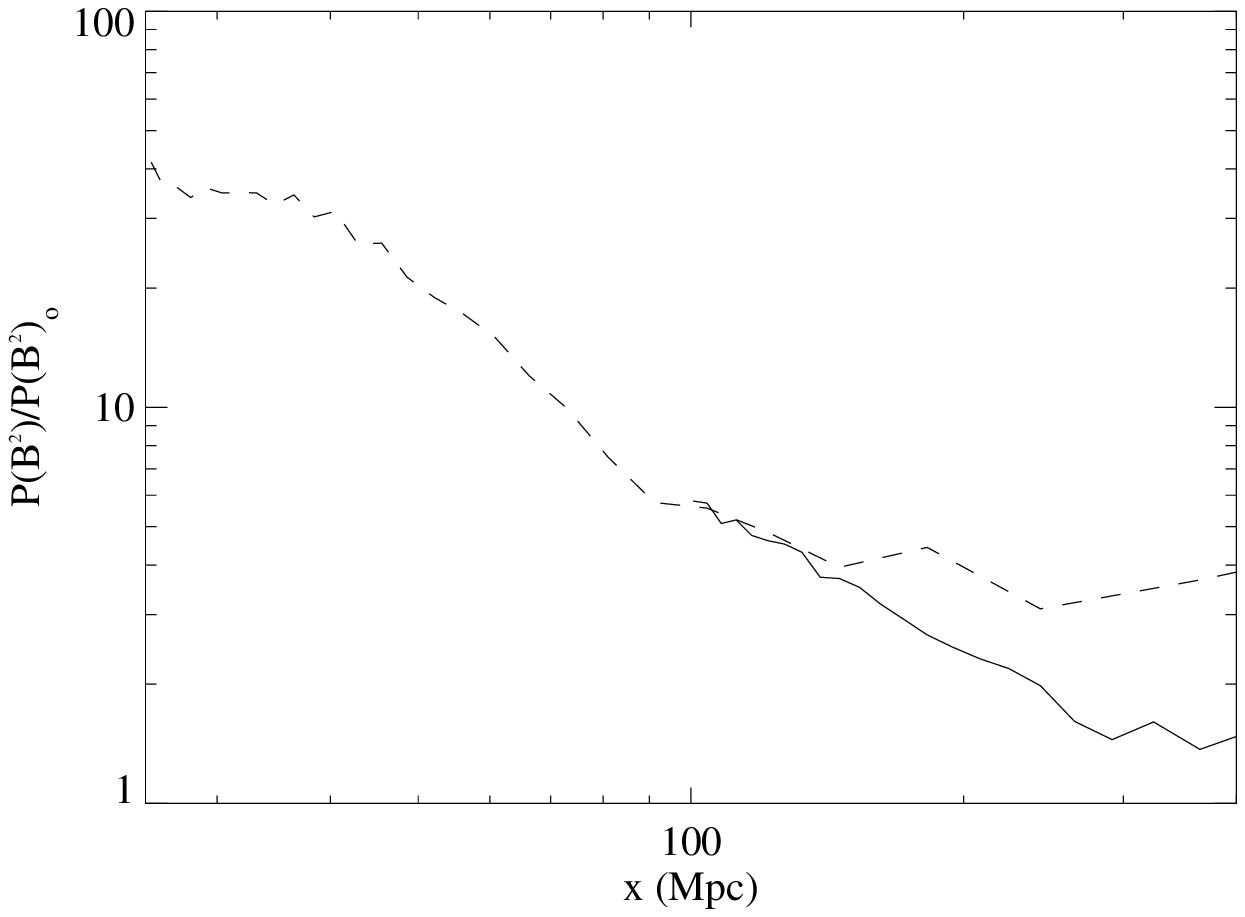}
\caption[]
{ }
\label{bpower_res}
\end{figure}

{\bf Fig. \ref{bpower_res}} The ratio of pre- to post-merger magnetic energy power spectra for
low (50 kpc/zone; solid line) and high (12.5 kpc/zone; dashed line) resolution simulations at 3.4 Gyrs. The analysis
was performed on an 800 kpc cube centered on the gravitational potential minimum. Field amplification
is significantly less in the low resolution simulation, and there is no significant amplification
on resolved scales ($>$4-5 zones) in the low resolution simulation.

\begin{table}
\label{tab1}
\begin{center}
\begin{tabular}{c c c c c c c c c } 
\multicolumn{9}{c}{Table 1. Initial Cluster Parameters}\\ \hline \hline

 \multicolumn{1}{c}{Cluster} & 
 \multicolumn{1}{c}{$M_{tot}^1$} &
 \multicolumn{1}{c}{$T_e^2$} &
 \multicolumn{1}{c}{$\sigma_v^3$} &
 \multicolumn{1}{c}{$r_c^4$} &
 \multicolumn{1}{c}{$f_g^5$} &
 \multicolumn{1}{c}{$\beta^6$}&
 \multicolumn{1}{c}{$n_{eo}^7$} &
 \multicolumn{1}{c}{$v_{impact}^8$} \\

 \multicolumn{1}{c}{ ID } & 
 \multicolumn{1}{c}{ (10$^{14}$ M$_\odot$) } &
 \multicolumn{1}{c}{ (keV) } &
 \multicolumn{1}{c}{ (km s$^{-1}$) } &
 \multicolumn{1}{c}{ (kpc) } &
 \multicolumn{1}{c}{     } &
 \multicolumn{1}{c}{     } &
 \multicolumn{1}{c}{ (10$^{-3}$cm$^{-3}$) } &
 \multicolumn{1}{c}{ (km s$^{-1}$)}  \\ \hline

Primary & 8.0 & 6.7 & 785  & 220 & 0.12 & 0.75 & 1.55 & 2300 \\
Subcluster & 3.2 & 3.1 & 526  & 135 & 0.06 & 0.72 & 1.01 &  \\ \hline 
\end{tabular}
\end{center}
{ $^1$ Total Mass R$<$3 Mpc.
 $^2$ Temperature.
 $^3$ Velocity Dispersion. \\
 $^4$ Core Radius. 
 $^5$ Global Gas Fraction, by mass. 
 $^6$ $\beta=\mu m_h \sigma^2/k T$;  R$<$1.5 Mpc.\\
 $^7$ Central Gas Density. 
 $^8$ Impact Velocity (dark matter).}
\end{table}

\newpage
\pagestyle{empty}
\onecolumn

\onecolumn

\end{document}